\documentclass[letter,twocolumn]{jpsj3}

\usepackage{bm}
\usepackage{graphicx}
\usepackage{amsmath}

\DeclareMathOperator{\sgn}{sgn}
\DeclareMathOperator{\sech}{sech}
\renewcommand{\Re}{\mathop{\mathrm{Re}}\nolimits}

\title{Thermoelectric Transport Coefficients for Massless Dirac Electrons in 
Quantum Limit}
\author{ \name{Igor \surname{Proskurin}}$^{1, 2}$\thanks{E-mail: iprosk@hosi.phys.s.u-tokyo.ac.jp}
and \name{Masao \surname{Ogata}}$^{1}$ }
\inst{$^{1}$ Department of Physics, University of Tokyo, 7-3-1 Hongo, Bunkyo-ku,
Tokyo 113-0033, Japan\\
$^{2}$ Department of Theoretical Physics, Institute of Natural Sciences, 
Ural Federal University, Mira 19, Ekaterinburg 620002, Russia}
\abst{We perform an analytic calculation of thermoelectric transport coefficients  for 
massless Dirac electrons using the approach based on the Kubo--St\v{r}eda formula 
and generalized Mott's relation. The main focus of the letter is made on the 
properties of the Nernst coefficient in the vicinity of the Dirac point 
in quantum limit. We calculate magnetic field and temperature dependencies of the Nernst coefficient
and compare our results with recent experiments in $\alpha$-(BEDT-TTF)$_{2}$I$_{3}$  
organic conductor. We argue that the Zeeman splitting is important to understand
the experimental data at high magnetic fields.}
\kword{Dirac electrons, Dirac point, thermoelectric coefficients, graphene,  
$\alpha$-(BEDT-TTF)$_{2}$I$_{3}$, Nernst effect, generalized Mott's formula}

\begin{document}
\maketitle

Unusual thermoelectric properties of graphene have attracted considerable interest. 
In graphene conducting electrons can be described by a Weyl equation that in quantizing
magnetic field leads to relativistic Landau levels with energies
$\pm \hbar\omega_{c}\sqrt{n}$ ($n=0,1,\ldots$) 
with $\omega_{c}$ and $\hbar$ being the cyclotron frequency and Plank's constant respectively.
The important difference of relativistic Landau levels from the non-relativistic case 
is the existence of $n=0$ level with zero energy.
In quantizing magnetic field, when the chemical potential is close to $n=0$ Landau level,
Seebeck and Nernst coefficients show an anomalous behaviour. \cite{Zuev2009,Checkelsky2009}

Massless Dirac fermions were also experimentally found in $\alpha$-(BEDT-TTF)$_{2}$I$_{3}$
organic conductor under pressure. \cite{Tajima2006,Fukuyama2012} 
Tight binding model  \cite{Katayama2006b,Kobayashi2007} 
and band structure calculations \cite{Ishibashi2006,Kino2006}
revealed the existence of the Dirac point in this material with conducting electrons
obeying the tilted Weyl equation.
Recently, an anomalously large Nernst signal was reported in this system 
at high magnetic fields. \cite{Konoike2012}.

Recently, the Nernst effect was intensively studied for massless Dirac
fermions in graphene \cite{Gusynin2006,Zhu2010,Ugarte2011} 
and on the surface of a topological insulator. \cite{Yokoyama2011}.
Various theoretical approaches showed that the Nernst coefficient is greatly 
enhanced and can considerably exceed the Seebeck coefficient when the chemical 
potential is close to the Dirac point \cite{Gusynin2006,Zhu2010,Ugarte2011}. 
This behaviour is sharply contrasted from the case with finite chemical potential 
in which the behaviour of the the transport 
coefficients is consistent with previous theoretical predictions
for the non-relativistic two-dimensional electron gas. \cite{Girvin1982,Jonson1984}

From a theoretical viewpoint the calculation of the Nernst coefficient
is a rather challenging problem since the standard linear response approach, 
based on the Kubo formula, gives unphysical divergence at zero temperature. 
Corrections arising from the
thermal magnetization should be taken into account. \cite{Smrcka1977,Cooper1997} 
Now it is well established that for the case of non-interacting electrons the 
thermopower tensor satisfies the generalized Mott's relation \cite{Smrcka1977,Cooper1997}
\begin{equation} \label{mott}
\bm{S} = \frac{\bm{\sigma}^{-1}(T,\mu)}{eT} \int_{-\infty}^{\infty} d \epsilon
f'\left( \epsilon \right) \left( \epsilon - \mu \right) \bm{\sigma}(0,\epsilon).
\end{equation}
where $T$ is temperature, $e>0$ is an electron charge, 
$\mu$ is chemical potential, $\bm{\sigma}$ is conductivity 
tensor, $f(\epsilon)=\left\{ 1 + \exp\left[\left( \epsilon-\mu \right)/k_{B}T \right]\right\}^{-1}$
is Fermi--Dirac distribution function, 
$f^{'}$ denotes the derivative with respect to $\epsilon$,
and $k_{B}$ is Boltzmann constant. The important property
of the Mott's formula is that, to calculate the thermopower tensor, one only needs
to know the conductivity at $T=0$ as a function of $\mu$,
$\bm{\sigma}(0,\mu)$.

The present letter is mainly devoted to the properties of the Nernst coefficient
of massless Dirac fermions in the vicinity of the Dirac point in quantum limit
where the distance between Landau levels is greater than temperature and impurity
broadening. For this purpose we perform an analytic 
calculation of thermoelectric coefficients using the Kubo--St\v{r}eda formula for conductivity and
generalized Mott's relation. We ignore the possible tilting
of the Dirac cone and consider a simple case of energy-independent damping $\Gamma$
due to the impurity scattering. 
In the limiting case when $\Gamma$ is much less than $k_{B}T$ and $\hbar\omega_{c}$, 
we obtain an analytical expressions for Seebeck and Nernst coefficients.
We show that the magnetic field dependence 
of the Nernst coefficient in the presence of the Zeeman splitting 
is different in  $\Gamma \ll k_{B}T$ and $\Gamma \geqslant k_{B}T$ regimes. 

In order to obtain the conductivity, $\bm{\sigma}(T,\mu)$, we consider a system of free massless 
Dirac electrons confined to a two-dimensional $\left( x,y \right)$-plane moving 
in a magnetic field $B$ perpendicular to the plane. The model Hamiltonian is given by
\begin{equation} \label{ham}
H = -v_{F} \sum_{i=x,y} \sigma_{i} \left[
-i \hbar \partial_{i} + e A_{i}(\bm{r})
\right]
\end{equation}
where $v_{F}$ is Fermi velocity, $\sigma_{i}$ is Pauli matrix,
$\partial_{i}$ denotes a derivative with respect to $i=x,y$, and
$\bm{A}(\bm{r})=\left( -By,0,0 \right)$ is a magnetic vector potential in the Landau gauge.

The solution of the eigenvalue problem for the Hamiltonian (\ref{ham}) leads to relativistic
Landau levels $E_{n\alpha} = \alpha \hbar \omega_{c} \sqrt{n}$ ($n=0,1,\ldots$) where
$\alpha = \pm 1$ is the band index and the cyclotron frequency is given by
$\omega_{c} = \sqrt{2} v_{F}/l_{B}$ with $l_{B}=\sqrt{\hbar/eB}$ being the magnetic length. 
The corresponding eigenfunctions are
\begin{equation} \label{psi0}
\psi_{k0}(x,y) = \frac{e^{ikx}}{\sqrt{l_{B}L}} \left(
\begin{array}{c}
0 \\ 
\phi_{0}\left( \frac{y}{l_{B}} - kl_{B} \right)
\end{array} \right)
\end{equation}
and 
\begin{equation} \label{psin}
\psi_{kn\alpha}(x,y) = \frac{e^{ikx}}{\sqrt{2l_{B}L}} \left(
\begin{array}{c}
\phi_{n-1}\left(\frac{y}{l_{B}} - kl_{B} \right) \\
\alpha \phi_{n}\left( \frac{y}{l_{B}} - kl_{B} \right)
\end{array} \right)
\end{equation}
for $n=1,2,\ldots$, where $\phi_{n}$ are the eigenfunctions of a harmonic oscillator.
We imply periodic boundary conditions in the $x$-direction with $L$ and $k$ being
the system size in the $x$-direction and the wave number respectively. 

For non-interacting electrons the conductivity can be calculated using the 
Kubo--St\v{r}eda formula \cite{Streda1975}
\begin{multline}
\sigma_{ij} = \frac{ie^{2}\hbar}{2\pi} \int_{-\infty}^{\infty} d\epsilon f(\epsilon) \\
\times \mathrm{Tr} \left[ v_{i} \frac{d G^{+}(\epsilon)}{d \epsilon} v_{j} A(\epsilon)
-v_{i}A(\epsilon) v_{j} \frac{d G^{-}(\epsilon)}{d\epsilon} \right]
\end{multline}
where Green functions are defined by 
$G^{\pm}(\epsilon)=\left( \epsilon - H \pm i\delta \right)^{-1}$,
$A = i(G^{+}-G^{-})$, and  
$\bm{v} = (i/\hbar) \left[ H,\bm{r}\right]$ is the velocity operator. Calculating the trace
with eigenfunction (\ref{psi0}) and (\ref{psin}) we obtain the following expressions for the
diagonal and off-diagonal parts of the conductivity
\begin{equation} \label{sxx}
\sigma_{xx} = - \frac{e^{3} v_{F}^{2} B}{16 \pi^{2}}
\sum_{\alpha\alpha'} \sum_{n=0}^{\infty} \int_{-\infty}^{\infty} d \epsilon
f'(\epsilon) A_{n+1\alpha}(\epsilon) A_{n\alpha'}(\epsilon), 
\end{equation}
\begin{multline} \label{sxy}
\sigma_{xy} = \frac{e^{3} v_{F}^{2} B}{8 \pi^{2}}
\sum_{\alpha\alpha'} \sum_{n=0}^{\infty} \int_{-\infty}^{\infty} d \epsilon
f(\epsilon) \\ \!\!\! \times \left[ \frac{d \Re G_{n+1\alpha}(\epsilon)}{d \epsilon}
A_{n\alpha'}(\epsilon)
-A_{n+1\alpha}(\epsilon)\frac{d \Re G_{n\alpha'}(\epsilon)}{d \epsilon} \right]
\end{multline}
where
\begin{equation}
\Re G_{n\alpha}(\epsilon) = \frac{\epsilon - E_{n\alpha}}
{\left(\epsilon - E_{n\alpha}\right)^{2}+\Gamma^{2}},
\end{equation}
\begin{equation} \label{a}
A_{n\alpha}(\epsilon) = \frac{2\Gamma}{\left(\epsilon - E_{n\alpha}\right)^{2}+\Gamma^{2}}.
\end{equation}
Here, in order to take into account the impurity scattering, we introduce a damping parameter $\Gamma$.

In the low field limit $\hbar \omega_{c} \ll k_{B}T$, we can replace the summation over Landau levels in 
Eqs.~(\ref{sxx}) and (\ref{sxy}) by an integration over continuous variable $E$. 
After performing the integration over $E$ and using the Mott's relation (\ref{mott}), 
in the leading order in magnetic field, we obtain longitudinal and transversal components
of the thermopower in terms of universal functions of $\hbar\omega_{c}/\Gamma$,
$k_{B}T/\Gamma$, and $\mu/\Gamma$
\begin{equation} 
S_{xx} = -\frac{k_{B}}{e} \frac{\tilde{K}^{0}_{xx}}{K^{0}_{xx}},
\end{equation}
\begin{equation} \label{sxylow}
S_{xy} = \frac{k_{B}}{e} \left(\frac{\hbar\omega_{c}}{2\Gamma}\right)^{2}
\frac{K^{0}_{xx} \tilde{K}^{0}_{xy}-K^{0}_{xy}\tilde{K}^{0}_{xx}}{\left(K^{0}_{xx}\right)^{2}}
\end{equation}
where
\begin{equation}
K^{0}_{ij} = \int_{-\infty}^{\infty} \frac{d x}{\cosh^{2} \frac{1}{2}x} 
\Phi_{ij}\left( \frac{k_{B}T}{\Gamma} x +\frac{\mu}{\Gamma} \right),
\end{equation}
\begin{equation} \label{tklow}
\tilde{K}^{0}_{ij} = \int_{-\infty}^{\infty} dx \frac{x}{\cosh^{2} \frac{1}{2}x} 
\Phi_{ij}\left( \frac{k_{B}T}{\Gamma} x +\frac{\mu}{\Gamma} \right).
\end{equation}
These formulae are in an agreement with the results obtained previously
using slightly different approaches. \cite{Gusynin2006,Ugarte2011}
The universal functions $\Phi_{ij}$ are the same as obtained previously for the case of
longitudinal and Hall conductivities calculations
\cite{Shon1998,Katayama2006a,Fukuyama2007}
\begin{equation} \label{phixx}
\Phi_{xx}(x) = 1+\left( x+ \frac{1}{x} \right) \tan^{-1} x,
\end{equation}
\begin{equation} \label{phixy}
\Phi_{xy}(x) = \frac{1}{x} \left( \frac{8x^{2}}{3(1+x^{2})^{2}} + 
\frac{1+x^{2}}{x} \tan^{-1} x -\frac{1-x^{2}}{1+x^{2}} \right).
\end{equation}
From the Eqs.~(\ref{phixx}) and (\ref{phixy}), using the expansion
$\Phi_{xx}\approx 2$, $\Phi_{xy} \approx 16x/3$ ($x \ll 1$) and
$\Phi_{xx}\approx (\pi/2)|x|$, $\Phi_{xy} \approx (\pi/2)\, \sgn x  $ ($x \gg 1$),
the following asymptotic behaviour for the Nernst coefficient 
at $\mu = 0$ can be obtained: \cite{Gusynin2006,Ugarte2011}
\begin{equation} \label{smallt}
S_{xy} = 
\frac{2\pi^{2}k_{B}^{2}T (\hbar \omega_{c})^{2}}{9 e\Gamma^{3}}, 
\quad \mbox{for $k_{B}T \ll \Gamma$.}
\end{equation}
and
\begin{equation} \label{hight}
S_{xy} = \frac{(\hbar\omega_{c})^{2}}{4e\Gamma T}, \quad \mbox{for $k_{B}T \gg \Gamma$.}
\end{equation}

On the other hand, in the quantum limit where Landau levels are well separated 
$\hbar\omega_{c} \gg \mathrm{max}\{k_{B}T,\Gamma\}$, one needs to
evaluate Eqs.~(\ref{sxx}) and (\ref{sxy}) numerically, except for the case
when $\Gamma \ll k_{B}T$.
In this case one can approximate the Lorentzian in Eq.~(\ref{a}) 
by a $\delta$-function.
Applying this approximation to Eqs.~(\ref{sxx}) and (\ref{sxy})
and using the Mott's formula (\ref{mott}) we obtain the following analytic results for  
thermoelectric coefficients
\begin{equation} \label{sxx1}
S_{xx} = -\frac{k_{B}}{e \vphantom{\left(\Gamma/k_{B}T\right)^{2}} } 
\frac{K_{xy} \tilde{K}_{xy} - \left(\Gamma/k_{B}T\right)^{2}
K_{xx}\tilde{K}_{xx}}{K_{xy}^{2}+ \left(\Gamma/k_{B}T\right)^{2} K_{xx}^{2}},
\end{equation}
\begin{equation} \label{sxy1}
S_{xy} = \frac{\Gamma}{eT\vphantom{\left(\Gamma/k_{B}T\right)^{2}}}
\frac{ K_{xx} \tilde{K}_{xy} + K_{xy}\tilde{K}_{xx}}
{  K_{xy}^{2}+ \left(\Gamma/k_{B}T\right)^{2} K_{xx}^{2}}
\end{equation}
where functions of $\varepsilon_{n\alpha}= E_{n\alpha}/(2k_{B}T)$ and 
$\mathrm{x} = \mu/(2k_{B}T)$ are introduced
\begin{equation} \label{kxx}
K_{xx} = \frac{1}{4} \sech^{2} \mathrm{x} +
\frac{1}{2} \sum_{n=1}^{\infty} \sum_{\alpha}
n \sech^{2}\left( \mathrm{x}-\varepsilon_{n\alpha} \right),
\end{equation}
\begin{equation}
\tilde{K}_{xx} = \frac{1}{2} \frac{\mathrm{x}}{\cosh^{2} \mathrm{x}}+
\sum_{n=1}^{\infty} \sum_{\alpha}
\frac{n \left( \mathrm{x} - \varepsilon_{n\alpha} \right) }
{\cosh^{2}\left( \mathrm{x} - \varepsilon_{n\alpha} \right)},
\end{equation}
\begin{equation}
K_{xy} = \frac{1}{2} \tanh \mathrm{x} + \frac{1}{4} \sum_{n=1}^{\infty} \sum_{\alpha}
\frac{\sinh 2\mathrm{x}}{\cosh \left( \mathrm{x} - \varepsilon_{n\alpha} \right)
\cosh \left( \mathrm{x} + \varepsilon_{n\alpha} \right)},
\end{equation}
\begin{equation} \label{dxy}
\tilde{K}_{xy} = \varphi \left( \mathrm{x} \right)+ \sum_{n=1}^{\infty} 
\sum_{\alpha} \varphi\left( \mathrm{x} - \varepsilon_{n\alpha} \right),
\end{equation}
and $\varphi(z) = \log \left( 2 \cosh z \right) - z \tanh z$. 
The corrections to $K_{xy}$ due to the impurity 
scattering are of order $ \Gamma/\left(\hbar \omega_{c}\right)$, \cite{Zheng2002}
and can be omitted.

In is important that, Eqs.~(\ref{sxx1}) and (\ref{sxy1}) interpolate two typical cases: (I)
when $\mu$ is away from the Dirac point and (II) when $\mu$ is close to the Dirac point 
($\mu \approx 0$). In the former case, one can safely neglect 
the terms of order $(\Gamma/k_{B}T)^{2}$, and obtain the 
results similar to the case of non-relativistic two-dimensional electron gas 
\cite{Girvin1982, Jonson1984}
\begin{equation}
S_{xx} = -\frac{k_{B}}{e} \frac{\tilde{K}_{xy} }{K_{xy}},
\end{equation}
\begin{equation}
S_{xy} = \frac{\Gamma}{e T}
\frac{K_{xx} \tilde{K}_{xy} + K_{xy}\tilde{K}_{xx}}{K_{xy}^{2}}.
\end{equation}
Here the thermopower has a sequence of peaks near the Landau levels. 
At low temperatures at each Landau level $S_{xx}$ has a universal value 
$- \sgn n(k_{B}/e)\,\log 2/n$, while
the Nernst coefficient in this region is small in comparison with $S_{xx}$.

In the latter case ($\mu \approx 0$), the behaviour of the 
thermoelectric coefficients changes significantly since $K_{xy}$ vanishes in
the vicinity of $n=0$ Landau level. 
As a result, the Nernst coefficient has a large peak at $\mu=0$ with the value
given by 
\begin{equation} \label{sxycl}
S_{xy} = \frac{k_{B}^{2}T}{e\Gamma}
\frac{\tilde{K}_{xy}}{K_{xx}}.
\end{equation}
At low $T$ the peak saturates at the value $4k_{B}^{2} T \log 2/ (e\Gamma)$, 
while $S_{xx}$ vanishes. 
In this region, for non-zero but small $\mu$, the Nernst coefficient can considerably exceed 
the Seebeck coefficient. 

\begin{figure}
\centerline{\includegraphics[scale=.35]{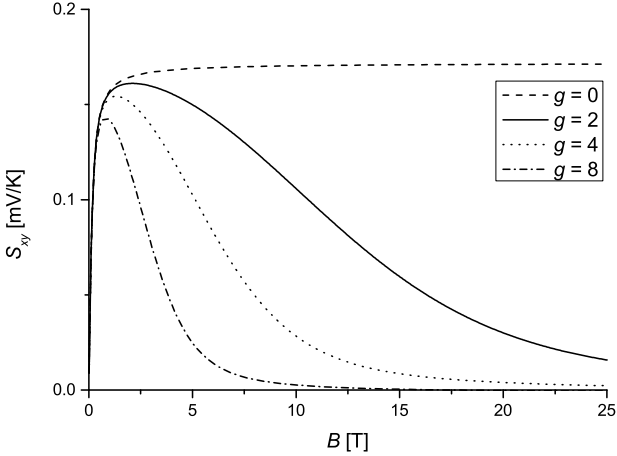}}
\caption{Magnetic field dependence of the Nernst coefficient at $\mu=0$ for $T=1.5$~K,
$v_{F} = 0.5 \times 10^{5}$~m/s, and $\Gamma/k_{B} = 3.75$~K for different
values of $g$-factor.}
\label{fig1}
\end{figure}

In order to describe the behaviour of the thermoelectric coefficients in high magnetic fields 
we also take into account Zeeman splitting of Landau levels 
$E_{n\alpha} \to E_{n\alpha} \pm \Delta $ where $\Delta = (1/2)g\mu_{B} B$, $\mu_{B}$  and $g$ are 
the Bohr magneton and $g$-factor respectively.
In magnetic field up to 10~T the Zeeman splitting is small
compared with $\hbar\omega_{c}$. For $v_{F}=0.5 \times 10^{5}$~m/s, $g = 2$, and
$B = 1$~T the ratio $\Delta/\hbar\omega_{c} \approx 0.03$.
The resulting magnetic field dependencies of the Nernst coefficient at $\mu=0$ for $T=1.5$~K 
and $v_{F}=0.5 \times 10^{5}$~m/s, and $\Gamma/k_{B}=3.75$~K 
for different values of $g$ are shown in Fig.~\ref{fig1}.
To obtain Fig.~\ref{fig1} we perform the numerical summation over
Landau levels in Eqs.~(\ref{sxx}), (\ref{sxy}) and use the Mott's formula (\ref{mott}).
The value $\Gamma/k_{B}=3.75$~K is chosen similar to that evaluated previously
from the magnetoresistance calculations for $\alpha$-(BEDT-TTF)$_{2}$I$_{3}$ .~\cite{Morinari2010}.
In high magnetic field $\hbar \omega_{c} \gg k_{B}T$, the Nernst coefficient in the case
without Zeeman splitting ($g = 0$) saturates as qualitatively predicted 
by Eq.~(\ref{sxycl}).
This saturating behaviour changes to a decay when Zeeman splitting becomes the same order as temperature 
and impurity broadening. In low magnetic field the asymptotic behaviour of $S_{xy}$ 
is described by Eqs.~(\ref{smallt}) and (\ref{hight}).

\begin{figure}
\centerline{\includegraphics[scale=.35]{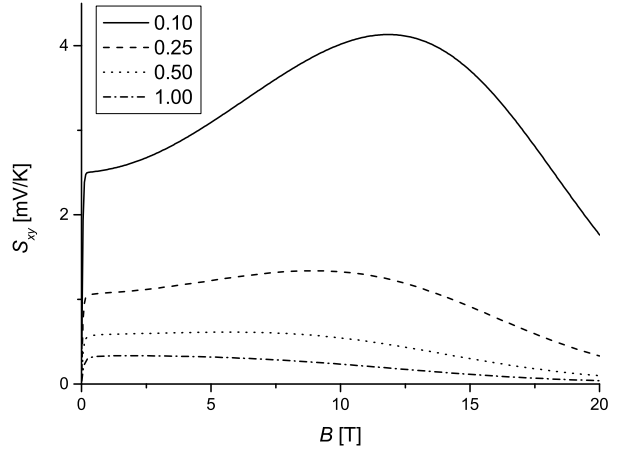}}
\caption{Magnetic field dependence of the Nernst coefficient at $\mu=0$ for $T=1.5$~K,
$v_{F} = 0.5 \times 10^{5}$~m/s, and $g=2$ for different
values of $\Gamma/(k_{B}T)$.}
\label{fig2}
\end{figure}

\begin{figure}
\centerline{\includegraphics[scale=.35]{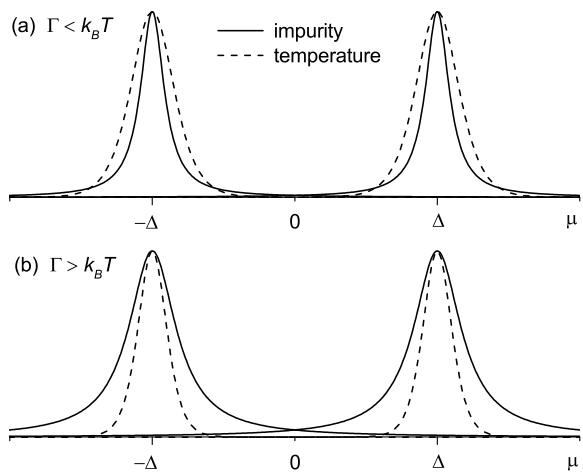}}
\caption{The difference between the Landau level broadening in (a) $\Gamma < k_{B}T$ and (b) $\Gamma > 
k_{B}T$ case. The impurity (solid) and temperature (dashed) broadenings are given by Lorentzian 
$\Gamma^{2}/\left[(\mu \pm \Delta)^{2}+\Gamma^{2}\right]$ and 
$\sech^{2}\left[(\mu \pm \Delta)/2k_{B}T\right]$ respectively.}
\label{fig4}
\end{figure}

The magnetic field dependence of the Nernst coefficient in the presence of Zeeman splitting
is different in the (a) $\Gamma \ll k_{B}T$ and (b) $\Gamma \geqslant k_{B}T$ limits 
which reflects the different mechanisms of Landau level broadening.
Figure~\ref{fig2} shows the magnetic field dependence of the Nernst coefficient for
several values of $\Gamma/(k_{B}T)$. In the case (b) with large $\Gamma/(k_{B}T)$
(dot-dash line in Fig.~\ref{fig2}), $S_{xy}$ decreases 
monotonically except for the very vicinity of $B=0$. In contrast, 
in the case (a) with small
$\Gamma/(k_{B}T)$ (solid line in Fig.~\ref{fig2}), 
there is a region in which $S_{xy}$ increases. 
In this case the increase the Nernst coefficient is understood from 
Eqs.~(\ref{kxx}), (\ref{dxy}), and (\ref{sxycl}).
Actually, the asymptotic behaviour for large $\Delta/(k_{B}T)$ is given by
\begin{multline}
S_{xy} = \frac{k_{B}^{2}T}{e\Gamma} 
\left[1 + \frac{\Delta}{k_{B}T} +\left(\frac{3}{2}+\frac{\Delta}{k_{B}T}\right)
e^{-\Delta/(k_{B}T)}\right]
\\ + O(e^{-2\Delta/(k_{B}T)}).
\end{multline}
However, the Nernst coefficient  starts to decrease in the large $B$ region.
This is understood as follows. 
For sufficiently large $\Delta$ the impurity broadening at $\mu = 0$, given by Lorentzian,
becomes dominant over the temperature broadening which has exponential decay, 
as illustrated in Fig.~\ref{fig4}~(a).
This effect causes the decay of the Nernst coefficient in the large $B$ region 
where the contribution at $\mu = 0$ comes from the overlap of the split 
$n = 0$ Landau level.
In case (b), the impurity broadening is always dominant over temperature broadening,
as illustrated in Fig.~\ref{fig4}~(b), and
behaviour of the Nernst coefficient is similar to that shown in Fig.~\ref{fig1}
and by dash-dot line in Fig.~\ref{fig2}.

\begin{figure}
\centerline{\includegraphics[scale=.35]{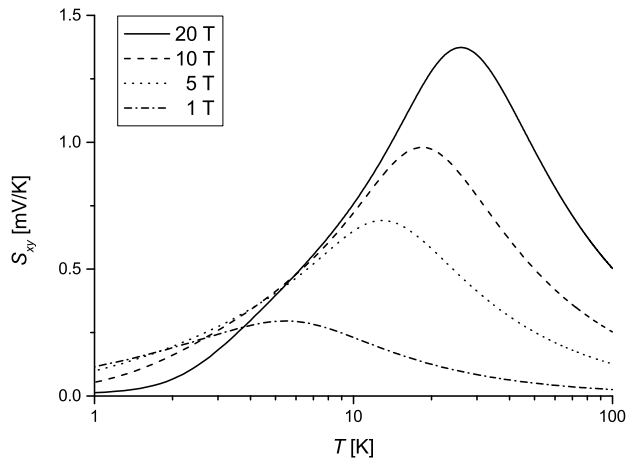}}
\caption{Temperature dependence of the Nernst coefficient at $\mu=0$ for
$v_{F} = 0.5 \times 10^{5}$~m/s and $\Gamma/k_{B} = 3.75$~K for different values
of magnetic field $B$.}
\label{fig3}
\end{figure}

The temperature dependence of the Nernst coefficient at $\mu=0$
for $v_{F}=0.5 \times 10^{5}$~m/s, $\Gamma/k_{B}=3.75$~K, and
$g=2$, calculated from Eqs.~(\ref{mott}), (\ref{sxx}) and(\ref{sxy}), 
is shown in Fig.~\ref{fig3} for several values of magnetic field. 
For large magnetic fields, $S_{xy}$ shows activation behaviour 
at low temperature due to the Zeeman splitting of $n=0$ Landau level,
while for small $B$ the temperature dependence at low temperature is approximately linear
as predicted by Eq.~(\ref{sxycl}). 
The position of the peak corresponds to the temperature when different Landau levels 
start to overlap which separates the quantum limit 
($\hbar \omega_{c} \gg k_{B}T$) from the low field limit ($\hbar \omega_{c} \ll k_{B}T$). 
In the latter case the asymptotic behaviour at high temperatures is given by
Eq.~(\ref{hight}).

In summary, we have calculated longitudinal and transverse components of the
thermopower in quantum limit. For the Nernst coefficient we have calculated the magnetic field
and temperature dependencies at $\mu = 0$. These results
with the Zeeman term are qualitatively consistent with the recent experiments 
in $\alpha$-(BEDT-TTF)$_{2}$I$_{3}$ organic conductor, \cite{Konoike2012} although there are some 
quantitative discrepancies. 
First, for the $g$-factor of $g=2$
the decay rate of $S_{xy}$ in Fig.~\ref{fig1} as a function of 
the magnetic field is about a factor 2 smaller  than that of experiment. 
To reproduce the experimental decay rate in our theory,
we need to assume  $g \approx 6$, which is similar to the effective g-factor 
discussed in Ref. \cite{Tajima2010} 
Second, the positions of the peaks on
the temperature dependencies shown in Fig.~\ref{fig3} are shifted to higher temperatures
than in the experiment. The origin
of this shift, as pointed out in Ref.~\cite{Sugawara2010}, may arise 
from magnetic field dependence of $\Gamma$ due to the presence 
of charged impurities. \cite{Shon1998,Zheng2002}
In order to achieve better agreement between the theory and experiments 
in $\alpha$-(BEDT-TTF)$_{2}$I$_{3}$, it will be necessary to take into account the tilting of the
Dirac cone and to use more realistic model for impurity scattering.
This remains as a future problem.

\begin{acknowledgements}
The authors are thankful to Takako Konoike and Hidetoshi Fukuyama for useful discussions.
This work is financially supported by Grant-in-Aid for Scientific Research (A)
(No.~24244053) from Japan Society for the Promotion of Science.
I.~P. acknowledges Russian Foundation for Basic Research, Grant No.~12-02-31565.
\end{acknowledgements}


\begin{thebibliography}{99}
\bibitem{Zuev2009} Y.~M.~Zuev, W.~Chang, P.~Kim: Phys. Rev. Lett. \textbf{102} (2009)
096807.
\bibitem{Checkelsky2009} J.~G.~Checkelsky, N.~P.~Ong: Phys. Rev.~B \textbf{80} (2009) 081413(R).
\bibitem{Tajima2006} N.~Tajima, S.~Sugawara, M.~Tamura, Y.~Nishio, K.~Kajita:
J.~Phys. Soc. Jpn. \textbf{75} (2006) 051010.
\bibitem{Fukuyama2012} H.~Fukuyama, Y.~Fuseya, M.~Ogata, A.~Kobayashi, Y.~Suzumura:
Physica~B \textbf{407} (2012) 1943.
\bibitem{Katayama2006b} A.~Katayama, A.~Kobayashi, Y.~Suzumura:
J.~Phys. Soc. Jpn. \textbf{75} (2006) 054705.
\bibitem{Kobayashi2007} A.~Kobayashi, S.~Katayama, Y.~Suzumura, H.~Fukuyama:
J.~Phys. Soc. Jpn. \textbf{76} (2007) 034711.
\bibitem{Ishibashi2006} S.~Ishibashi, T.~Tamura, M.~Kohyama, K.~Terakura:
J.~Phys. Soc. Jpn. \textbf{75} (2006) 015005.
\bibitem{Kino2006} H.~Kino, T.~Miyazaki: J.~Phys. Soc. Jpn. \textbf{75} (2006) 034704.
\bibitem{Konoike2012} T.~Konoike, K.~Uchida, M.~Sato, T.~Osada: presented at MDF2012, Int. 
Symposioum on Material Science Opened by Molec. Degrees of Freedom, 2012.
\bibitem{Gusynin2006} V.~P.~Gusynin, S.~G.~Sharapov: Phys. Rev.~B \textbf{73} (2006) 245411.
\bibitem{Zhu2010} L.~Zhu, R.~Ma, L.~Sheng, M.~Liu, D.-N.~Sheng: Phys. Rev. Lett. 
\textbf{104} (2010) 076804.
\bibitem{Ugarte2011} V.~Ugarte,V.~Aji, C.~M.~Varma: Phys. Rev.~B \textbf{84} (2011)165429.
\bibitem{Yokoyama2011} T.~Yokoyama, S.~Murakami: Phys. Rev.~B \textbf{83} (2011) 161407(R).
\bibitem{Girvin1982} S.~M.~Girvin, M.~Jonson: J.~Phys. C -- Solid State \textbf{15} (1982) L1147.
\bibitem{Jonson1984} M.~Jonson, S.~M.~Girvin: Phys. Rev.~B \textbf{29} (1984) 1939.
\bibitem{Smrcka1977} L.~Smr\v{c}ka, P.~St\v{r}eda: J.~Phys. C -- Solid State \textbf{10} (1977) 2153.
\bibitem{Cooper1997} N.~R.~Cooper, B.~I.~Halperin, I.~M.~Ruzin: Phys. Rev.~B \textbf{55}
(1997) 2344.
\bibitem{Streda1975} P.~St\v{r}eda, L.~Smr\v{c}ka: Phys. Status Solidi B \textbf{70} (1975) 537.
\bibitem{Shon1998} N.~H.~Shon, T.~Ando: J.~Phys. Soc. Jpn. \textbf{67} (1998) 2421.
\bibitem{Katayama2006a} S.~Katayama, A.~Kobayashi, Y.~Suzumura: J.~Phys. Soc. Jpn.
\textbf{75} (2006) 023708.
\bibitem{Fukuyama2007} H.~Fukuyama: J.~Phys. Soc. Jpn. \textbf{76} (2007) 043711.
\bibitem{Zheng2002} Y.~Zheng, T.~Ando: Phys. Rev.~B \textbf{65} (2002) 245420.
\bibitem{Morinari2010} T.~Morinari, T.~Tohyama: J.~Phys. Soc. Jpn. \textbf{79} (2010)
044708.
\bibitem{Tajima2010} N.~Tajima, M.~Sato, S.~Sugawara, R.~Kato, Y.~Nishio, K.~Kajita:
Phys. Rev.~B \textbf{82} (2010) 121420(R).
\bibitem{Sugawara2010} S.~Sugawara, M.~Tamura, N.~Tajima, R.~Kato, M.~Sato,
Y.~Nishio, K.~Kajita: J.~Phys. Soc. Jpn. \textbf{79} (2010) 113704.
\end{thebibliography}
\end{document}